\documentclass[nofootinbib,twocolumn,longbibliography]{revtex4-1}
\usepackage[english]{babel}
\usepackage[autostyle, english = british]{csquotes}
\usepackage[bookmarks=false]{hyperref}
\usepackage{amsmath,amssymb}
\usepackage{graphicx}
\usepackage{enumitem}
\usepackage{color}
\usepackage[T1]{fontenc}
\setlist[enumerate]{topsep=0pt,parsep=-1mm,leftmargin=5mm,}
 \MakeOuterQuote{"}
\usepackage{soul}

\usepackage[left=2.5cm,top=2cm,right=2.5cm,bottom=2cm]{geometry}

\begin{document}

\title{\large Cosmic inflation prevents singularity formation in collapse into a Hayward black hole}

\author{Micha{\l} Bobula${}^{a, b}$}
\email{michal.bobula@uwr.edu.pl}
\affiliation{${}^a$University of Wroc{\l}aw, Faculty of Physics and Astronomy, Institute of Theoretical Physics, pl. M. Borna 9, 50-204 Wroc{\l}aw, Poland}
\affiliation{${}^b$Instituto de Estructura de la Materia, IEM-CSIC, Serrano 121, 28006 Madrid, Spain}

\begin{abstract} 
\noindent We construct a (quantum mechanically) modified model for the Oppenheimer-Snyder collapse scenario where the exterior of the collapsing dust ball is a Hayward black hole spacetime and the interior is a dust Friedmann-Robertson-Walker cosmology. This interior cosmology is entirely determined by the junction conditions with the exterior black hole. It turns out to be non-singular, displaying a power-law contraction which precedes a de Sitter phase or, reversely, a power-law expansion followed by a de Sitter era. We demonstrate that cosmic inflation in the collapse setting is a mechanism that decelerates collapsing matter, thereby preventing singularity formation. We also analyse the global causal structure and the viability of the model. 
\end{abstract}

\maketitle

\section{Introduction}

The question of whether the spacetime singularities inherent to General Relativity (GR) represent a real physical phenomenon remains open more than a century after the advent of this gravitational theory. In cosmological scenarios, the debate affects fundamental questions about the origin of the Universe. Was there something before the Big Bang? Did the Universe have a beginning? In black hole physics, the problem is related to other important physical issues. What is the fate of free-falling observers? What happens beyond the event horizon? The classical Oppenheimer-Snyder (OS) collapse scenario \cite{Oppenheimer:1939ue} is a especially suitable arena to discuss the connection between the aforementioned questions. The fact that we can model a collapsing dust ball using a Friedmann-Robertson-Walker (FRW) metric in its interior and a Schwarzschild solution in the exterior suggests a link between black hole spacetimes and cosmology. Nevertheless, the resulting geometry still possesses a singularity, posing serious difficulties at the classical level and, more importantly, sustaining the black hole information paradox in a semiclassical picture of the black hole evaporation \cite{Ashtekar:2005cj}.

The singular nature of the Schwarzschild spacetime has motivated considerable effort devoted to exploring alternatives \cite{Poisson:1988wc, Dymnikova:1992ux,Hayward:2005gi, Maier:2009xp, Barragan:2009sq, Bambi:2013caa, Saini:2014qpa,DeLorenzo:2014pta,Frolov:2015bta,Frolov:2015bia,Frolov:2016pav,Carballo-Rubio:2018pmi,Kiefer:2019csi, Mosani:2020ena, Schmitz:2020vdr, Gozdz:2022dsa, Bonanno:2022jjp,Shafiee:2022jfx, Kiefer:2023zxt, Shojai:2022pdq, Malafarina:2022oka, Alonso-Bardaji:2023qgu, Duque:2023syb, Barca:2023shv, BobulaLOOPS,Bobula:2023kbo,Lewandowski:2022zce, Giesel:2022rxi, Giesel:2023hys, Giesel:2023tsj, Giesel:2024mps, Kelly:2020lec, Kelly:2020uwj, Husain:2021ojz, Husain:2022gwp, Fazzini:2023ova, Fazzini:2023scu, Cafaro:2024vrw, Cipriani:2024nhx, Han:2024rqb}. A prominent example is the Hayward black hole \cite{Hayward:2005gi}, on which we will focus our attention in this work. Instead of the singularity at the origin of the radial coordinate, this spacetime displays a regular "quantum gravity 
core". The Hayward geometry has a really special conformal diagram, representing an evaporating non-singular black hole (in which back-reaction is effectively introduced) \cite{Hayward:2005gi}\footnote{ Although the associated static metric was originally derived by Poisson \& Israel from the model inspired by one-loop corrections to Einstein field equations \cite{Poisson:1988wc}, its relevance for the information puzzle was established in Ref. \cite{Hayward:2005gi}.}. All matter entering the black hole ultimately reaches future null infinity. Hence, in a semiclassical approximation, most probably there is enough Cauchy data at this future infinity to reconstruct the past, resolving the information paradox.

Although the Hayward metric seems to solve some fundamental problems present in the Schwarzchild metric, there has been no consensus about which physical theory the metric could be derived from. However, recently the metric has been derived as a vacuum solution within two independent effective descriptions incorporating quantum gravity corrections \cite{Bueno:2024dgm, Giesel:2024mps}. The first one, called Quasi-Topological Theory \cite{Aguilar-Gutierrez:2023kfn,Bueno:2024dgm}, can be regarded as Einstein-Hilbert action supplemented with an infinite tower of higher-curvature corrections which are contractions of the Riemann tensor. For instance, when the tower is truncated at the second order, one gets Gauss-Bonnet gravity (see Refs. \cite{Arciniega:2018tnn,Bueno:2023dpl}). In general, such an infinite tower of corrections, weighted by powers of e.g. the Planck length, is what one could expect from String Theory (ST) as an ultraviolet completion of GR \cite{Bueno:2023dpl}. Specifically, when the number of spacetime dimensions $D$ is equal or greater than five and spherical symmetry is imposed, the equations of motions (being of second differential order) determine the Hayward metric as a unique vacuum solution, provided that the simplest sequence of gravitational couplings is chosen \cite{Bueno:2024dgm}. The other aforementioned description \cite{Giesel:2024mps}, built in four spacetime dimensions, leads to the Hayward metric as the vacuum solution coming from a deformed (or ``polymerized'') gravitational Hamiltonian in Lema\^{i}tre-Tolman-Bondi (LTB) spacetimes. Motivated by dust collapse models in Loop Quantum Gravity (LQG), where the effective dynamics can be regarded as an LTB model with a polymerized Hamiltonian \cite{Giesel:2023hys, Giesel:2023tsj, Giesel:2024mps}, the authors of Ref. \cite{Giesel:2024mps} construct a specific ``polymerization function'' for the Hamiltonian that reproduces the Hayward metric in vacuo. Contrary to what happens in standard effective LQG models, the polymerization function is now unbounded. Moreover, this description can be translated into a Lagrangian mimetic gravity theory with a concrete mimetic potential.

In the present work, we will assume that the Hayward metric describes the vacuum exterior of a modified OS collapse scenario. We will then impose conditions for the interior that we expect are compatible with the above descriptions and derive the corresponding cosmological behavior. Indeed, we will explore the resulting interior dust cosmology when we require: a) both the exterior and interior metrics join smoothly at the boundary, located at the surface of the collapsing dust ball, b) a conservation law for dust is satisfied. This investigation should open the door to exploring how well the interior FRW background describes our universe.

The interior geometry will be completely determined by imposing the Israel-Darmois junction conditions at the boundary. Within GR, Israel-Darmois junction conditions require induced metric and extrinsic curvature to be continuous at the junction surface provided no energy layer or thin shell being present there, which is the case for OS collapse scenario. We point out that these junction conditions were originally derived for General Relativity, thus, in general, when working with dynamical equations different from the Einstein Field Equations (EFEs), other conditions might be required \cite{Chu:2021uec}. However, based purely on geometric considerations --- without reference to any dynamical equations --- we observe that the Gauss-Codazzi equations (see Eqs. 3.39–3.40 in Ref. \cite{Poisson:2009pwt}) relate the extrinsic curvature to the full Riemann tensor evaluated on a hypersurface (the junction surface). Consequently, if we desire the full Riemann tensor to be continuous at the junction surface, the Gauss-Codazzi equations suggest that the extrinsic curvature must also be continuous on that surface (and vice versa). Therefore, we expect (but we do not prove) this geometric requirement --- namely, the continuity of the extrinsic curvature --- to hold in frameworks that identify the Hayward black hole as the unique vacuum solution in the context of Oppenheimer-Snyder collapse. In these descriptions, the connection is Levi-Civita, and the aforementioned Gauss-Codazzi equations remain valid. Indeed, since these frameworks do not introduce additional matter fields or degrees of freedom beyond those in GR, we expect that the requirement of metric and extrinsic curvature continuity at the surface of the dust ball (junction surface) is compatible with them. As we will show, merely requiring continuities of metric and extrinsic curvatures (Israel-Darmois junction conditions) fully determines the interior FRW geometry. However, we do not rigorously derive junction conditions viable for Quasi-Topological gravity and effective LQG-inspired frameworks. As shown in Ref. \cite{Chu:2021uec} for certain theories beyond GR, imposing Israel-Darmois junction conditions is inconsistent with a vanishing surface energy-momentum tensor or additional, nontrivial relations between the extrinsic and intrinsic curvatures have to be satisfied. For these reasons, we emphasize that the validity of the Israel-Darmois junction conditions remain an essential assumption of our work. To further support the application of the Israel-Darmois conditions we observe that i) in Gauss-Bonnet gravity (which can be regarded as Quasi-Topological gravity truncated at second order), the requirement of continuity of extrinsic curvature is viable (see, e.g., the discussion around Eqs. 3.67–3.68 in Ref. \cite{Chu:2021uec}) and ii) in the framework of LQG-inspired LTB models (however, with the standard polymerization function), it was demonstrated in Ref. [40] that applying the Israel-Darmois junction conditions and solving the dynamical equations yield the same results in the modified Oppenheimer-Snyder collapse scenario. As an aside, note that the formalism of Israel-Darmois junction conditions in the context of Oppenheimer-Snyder collapse has been successfully applied to models beyond the Einstein Field Equations by several authors in Refs. \cite{Maier:2009xp, Schmitz:2020vdr, Malafarina:2022oka, BobulaLOOPS, Lewandowski:2022zce, Bobula:2023kbo, Han:2023wxg, Giesel:2023tsj, Giesel:2023hys, Giesel:2024mps, Cafaro:2024vrw, Bonanno:2023rzk}.

The dust cosmology that we will obtain has the following remarkable features. It has no curvature singularities, it presents a de Sitter era preceding (following) a power-law expansion (contraction), and the transition between these two regimes is smooth, displaying a graceful exit. Moreover, we will see that our modified OS model shares the behavior found in Ref. \cite{Bonanno:2023rzk} for a model inspired by Asymptotically Safe Gravity (ASG). Finally, we will also construct and discuss the conformal diagram representing the modified OS collapse (without back-reaction due to Hawking radiation).

\section{Cosmological dynamics from junction conditions}

We consider a modified OS collapse scenario, where the static exterior of the collapsing dust ball is given by the line element
\begin{equation} \label{ext}
    \mathrm{d}s^2_{\mathrm{ext}} = -F(X) \mathrm{d}t^2 + \frac{1}{F(X)} \mathrm{d} X^2 + X^2 \mathrm{d}\Omega^2 \, ,
\end{equation}
where $\mathrm{d}\Omega^2 = \mathrm{d} \theta^2 + \sin ^2 \theta \mathrm{d} \varphi^2$ is the metric on the two-sphere, and the function $F(X)$ is chosen to correspond to a Hayward black hole \cite{Hayward:2005gi}, namely
\begin{equation} \label{Fhay}
F(X)=1-\frac{2 M X^2}{X^3+2 l^2 M} \, .
\end{equation}
Here, $M$ is a "mass" constant. For large values of the radial coordinate $X$ the metric behaves like the Schwarzchild one. The other constant in Eq. \eqref{Fhay} is usually taken to be the Planck length (and hence $l=1$ in the Planck units that we are adopting), although we will maintain it explicitly in our formulae. For small $X$, we have $F(X) \sim 1-X^2 / l^2$. Thus, we can expect that the cosmological interior that matches smoothly at the boundary displays an epoch of de Sitter behavior. Throughout the rest of this article, we restrict the analysis to the non-extreme case when $M>(3 \sqrt{3} / 4) l$ in Eq. \eqref{Fhay}, so that the mass ranges over values of astrophysical interest. In this case, $F(X)$ has two positive roots, say $0<X^-<X^+$. For the remaining cases $M \leq (3 \sqrt{3} / 4) l$ the causal structures will be different, however, the corresponding analysis is beyond the scope of this work.

For the homogeneous interior of the dust ball, on the other hand, we assume a flat FRW line element
\begin{equation} \label{intmetric}
\mathrm{d} s_{\mathrm{int}}^2=-\mathrm{d} T^2+a(T)^2 \mathrm{~d} r^2+r^2 a(T)^2 \mathrm{~d} \Omega^2.
\end{equation}
The form of the scale factor $a(T)$ is determined by our junction conditions with the exterior. In full detail, we demand that the Israel-Darmois junction conditions (see e.g. Chapters 3.7-3.8 of Ref. \cite{Poisson:2009pwt}) be satisfied at the surface of the collapsing dust ball. We call $\Sigma$ this boundary. Thus, the induced metric on $\Sigma$ and  its extrinsic curvature (second fundamental form) must coincide when computed from the exterior and from the interior, namely,  $h_{ab}^{\mathrm{ext}} = h_{ab}^{\mathrm{int}}$ and  $K_{ab}^{\mathrm{ext}} = K_{ab}^{\mathrm{int}}$, respectively. A detailed analysis of these conditions on the line elements \eqref{ext} and \eqref{intmetric} was recently carried out in Refs.  \cite{Bobula:2023kbo, Lewandowski:2022zce,Bobula:2024}\footnote{See Ref. \cite{Bobula:2024} for the general analysis of the case where the exterior line element \eqref{ext} is not necessarily stationary.}. Here, we will summarize the most important aspects. Let $T$ be a parameter at $\Sigma$. The trajectory of the free-falling observer at $\Sigma$ can be described either as $x^{\alpha}_{\mathrm{ext}} = (t=\tau(T), X=R(T)$) from the exterior or as $x^{\alpha}_{\mathrm{int}} = (T, r=r_b$) from the interior, where $r_b$ is a constant. The junction conditions form a soluble system of algebraic equations. They imply the following relations \cite{Bobula:2023kbo, Lewandowski:2022zce, Bobula:2024}, that are crucial for this work: 
\begin{equation} \label{j1}
    X\Big|_\Sigma = R(T) = r_b a(T) \, ,
\end{equation}
and
\begin{equation} \label{j2}
    F\Big|_\Sigma=1-\dot{R}(T)^2 \, ,
\end{equation}
where the dot denotes the derivative with respect to $T$. Note that $T$ is the affine parameter for observers co-moving with the surface. In order to arrive at the above two equations, no particular form of $a(T)$ or $F(X)$ was assumed.

We can combine Eqs. \eqref{Fhay}, \eqref{j1}, and \eqref{j2} to get
\begin{equation}
    1-\frac{2 M R(T)^2}{R(T)^3+2 l^2 M} = 1 - \dot{R}(T)^2 \, .
\end{equation}
Integrating the negative root of $\dot{R}(T)$ (so that matter is contracting) with the initial condition $R(0)=R_0=r_b a_0$, we obtain
\begin{equation} \label{Tofa}
\begin{split} 
&  T(a) = \frac{1}{3} \sqrt{\frac{2 a_{0}^3 r_{b}^3}{M}+4 l^2}-\frac{1}{3}\sqrt{\frac{2 a^3 r_{b}^3}{M}+4
   l^2} \\ &- \frac{1}{3} l \log \left(\frac{\left(\sqrt{\frac{2 a_{0}^3 r_{b}^3}{M}+4 l^2}+2 l\right) \left(\sqrt{\frac{2 a^3
   r_{b}^3}{M}+4 l^2}-2 l\right)}{\left(\sqrt{\frac{2 a_{0}^3 r_{b}^3}{M}+4 l^2}-2 l\right) \left(\sqrt{\frac{2 a^3
   r_{b}^3}{M}+4 l^2}+2 l\right)}\right) \, .
\end{split}
\end{equation}
It is challenging to invert this equation to get $a(T)$. However, we can analyse it to deduce important properties of the resulting scale factor. One has $\lim_{a\rightarrow \infty} T(a) = -\infty$ and $\lim_{a\rightarrow 0} T(a) = \infty$. These limits, together with the fact that relation \eqref{Tofa} is monotonic in $a \in (0, \infty)$, indicate that the inverse function $a(T)$ describes a timelike geodesically complete dust universe, in which matter is contracting for all values of the affine parameter $T\in (-\infty,\infty)$. Thanks to homogeneity, the scale factor obtained by solving the junction conditions is valid not only at $\Sigma$, but in the whole interior. Note that the $\log$ term dominates in Eq. \eqref{Tofa} for small $a$, implying that the contraction of the Universe is exponential in this regime. On the other hand, for large $a$, the $\log$ contribution becomes negligible and the scale factor approximately behaves as for dust collapse in GR, namely, $a(T) \sim T^{2/3}$. Since the dependence of $T(a)$ given in Eq. \eqref{Tofa} is smooth and strictly monotonic for $a\in (0, \infty)$, the deduced interior geometry admits a graceful entrance to the de Sitter epoch. It is worth remarking that the Kretschmann scalar $\mathcal{K}_{\mathrm{int}}$ is bounded everywhere in the interior. Indeed, this curvature invariant for the line element \eqref{intmetric}, with the scale factor $a$ determined by the inverse of $T(a)$ given in Eq. \eqref{Tofa}, can be calculated to be
\begin{equation}
\begin{split}
    \mathcal{K}_{\mathrm{int}} &= \frac{12 \left(a^2 T''(a)^2+T'(a)^2\right)}{a^4 T'(a)^6} = \\
&=\frac{12 M^2 \left(5 a^6 r_b^6+8 a^3 l^2 M r_b^3+32 l^4 M^2\right)}{\left(a^3 r_b^3+2  l^2 M\right)^4} \, . 
\end{split}
\end{equation}
In particular, in the limit $a\rightarrow 0$, when the metric \eqref{intmetric} becomes degenerate, we have $ \mathcal{K}_{\mathrm{int}} \rightarrow 24/l^4$.

If we write the energy density of the modified model of OS collapse as
\begin{equation} \label{density}
   \rho= \frac{3M}{4\pi r_b^3 a^3} \, ,
\end{equation}
a simple calculation leads us to the (quantum mechanically) modified Friedmann equations 
\begin{equation} \label{friedman}
    \left(\frac{\dot{a}}{a}\right)^2 = \frac{1}{\left(T'(a) \, a\right)^2 } = \frac{8 \pi  \rho }{3+ 8 \pi  l^2 \rho} \, ,
\end{equation}
\begin{equation}
    \dot{H}+H^2 = \frac{ 4 \pi  \rho  \left(-3+16 \pi  l^2 \rho \right)}{\left(3+8 \pi  l^2 \rho \right)^2} \, ,
\end{equation}

\begin{figure}
    \centering
    \includegraphics[width=7cm]{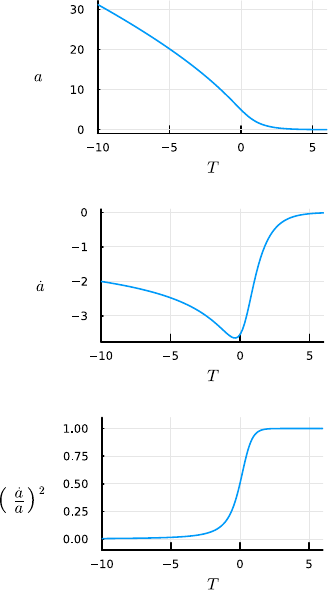}
    \caption{Time dependence of the scale factor, of its first time derivative, and of the square Hubble parameter (from top to bottom). The Universe follows a power-law contraction at early times and them smoothly transits to a de Sitter-like epoch at late times. We have taken $M=15$ and $a_0 =5$ (in Planck units).}
    \label{3wyk}
\end{figure}

where $H = \frac{\dot{a}}{a}$ and the prime stands for the derivative with respect to $a$. Condition \eqref{density} is the consequence of the conservation of the energy-momentum tensor for dust, $ \nabla_{\mu} \mathcal{T}^{\mu\nu}=0$, where  $\mathcal{T}^{\mu\nu}=\rho k^\mu k^\nu$ and $k^\mu \partial_\mu = \partial_T $. It is compatible with the two mentioned theories that identify the Hayward metric as a vacuum solution \cite{Bueno:2024dgm,Arciniega:2018tnn,Giesel:2024mps}\footnote{The conservation law indeed holds for the four-dimensional variant of the theory with an infinite tower of higher-order corrections when the FRW ansatz is adopted (see the discussion around Eq. (7) in Ref. \cite{Arciniega:2018tnn}).  }. The corrected Friedmann equations given above reproduce the dynamical equations of a dust FRW cosmology in GR both in the low energy limit (at leading order when $\rho$ is small) and in the limit $l\rightarrow 0$. Note also that, in the limit $a\rightarrow 0$, the energy density is clearly infinite. This issue will be discussed in Sec. \ref{SECdis}. 

\begin{figure*}
    \centering
    \includegraphics[width=\textwidth]{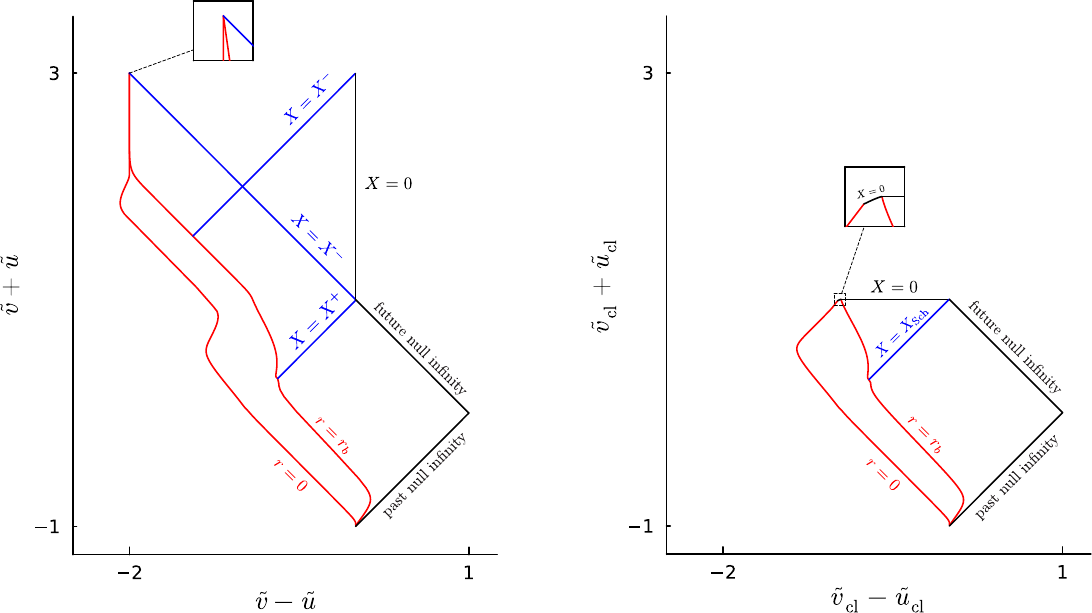}
    \caption{ \emph{Left:} the conformal diagram representing our (quantum mechanically) modified OS collapse scenario. The FRW geometry, enclosed by curves marked by $r=0$ and $r=r_b$, is surrounded by the static Hayward black hole geometry, with horizons $X=X_-$ and $X=X_+$. \emph{Right:} the conformal diagram for the classical OS collapse scenario\cite{Oppenheimer:1939ue}. Singular FRW geometry, whose dynamics is governed by the classical dust Friedmann equation $ \left(\frac{\dot{a}}{a} \right)^2 = \frac{8 \pi \rho}{3}$ is surrounded by Schwarzschild black hole geometry --- the metric function for the exterior is $F \rightarrow F_{\mathrm{Sch}}(X) = 1-\frac{2 M}{X}$ with Schwarzchild radius located at $X = X_{\mathrm{Schw}} = 2M$ and the singularity $X=0$. To generate both of the diagrams we have taken the same constants' values $M=15$, $a_0 =5$, $r_b=(4 \pi / 3)^{-1 / 3}$ and $\kappa=0.1$ (in Planck units). }
    \label{diagram}
\end{figure*}

We have numerically computed the scale factor, its first $T$-derivative, and the square Hubble parameter (for $M=15$ and $a_0 =5$ in Planck units). We display the results in Fig. \ref{3wyk}. The most crucial step in this computation is inverting Eq. \eqref{Tofa} to calculate $a(T)$. For this goal, we used libraries of \textsc{Julia} programming language. This numerical calculation supports our statements in the previous paragraphs, concerning the smooth transition from de Sitter to power-law expansion, providing a graceful exit. 

\section{Conformal Diagram}

In order to discuss the properties of the model, in particular in the context of the black hole information paradox, a useful tool is the conformal diagram of our spacetime. We have numerically generated the diagram in a rigorous manner that we comment in the following. This rigorous and unambiguous procedure is important to avoid confusions in the interpretation of the results, which are presented in Fig. \ref{diagram}.

We first adopt Painlev\'{e}-Gullstrand (PG) coordinates for the exterior. They simplify the construction, because the PG time and the time coordinate in Eq. \eqref{intmetric} coincide\footnote{This follows from the fact that the PG time coordinate is, by construction, the affine parameter of free-falling observers, exactly the same as the time coordinate used in Eq. \eqref{intmetric}.}. Thus, we can write $\mathrm{d}T=\mathrm{d}t + (1/F) \sqrt{1-F} \mathrm{d}X$, and metric \eqref{ext} becomes
\begin{equation}
    \mathrm{d}s^2_{\mathrm{ext}} = -\mathrm{d}T^2 + \left(\mathrm{d}X+\sqrt{1-F(X)}\mathrm{d}T\right)^2 + X^2\mathrm{d}\Omega \, .
\end{equation}
Next, let's introduce double null coordinates
\begin{equation} \label{doublenull}
    \begin{split}
        u= T - \int^X_0 \frac{1}{1-\sqrt{1-F}} \mathrm{d} \tilde{X} \, , \\
        v= T+ \int^X_0 \frac{1}{1+\sqrt{1-F}} \mathrm{d} \tilde{X} \, .
    \end{split}
\end{equation}

The integrand in the expression of $v(T,X)$ is regular, whereas the one for $u(T,X)$ is not. However, it can be handled with methods similar to those presented in Ref. \cite{Bobula:2023kbo} (see the discussion around Eqs. 2.29-3.31). These null coordinates still need to be appropriately compactified. To cover the exterior region in the left diagram in Fig. \ref{diagram}, we perform the compactification
\begin{equation} \label{comp}
    \begin{split}
      \tilde{u}&=\pm \arctan(\kappa \, u) /\pi + c_u \, , \\
        \tilde{v}&=\pm  \arctan(\kappa \, v) /\pi+ c_v \, ,
    \end{split}
\end{equation}
where $c_v \in \{0,1 \}$, $c_u \in \{0,1,2 \}$, and $\kappa$ are suitably chosen constants. For the sake of an example, consider the trajectory of the surface given by the pair $\{u(T,R(T)),v(T,R(T)\}$. The $u$ coordinate reaches infinity exactly at $R\rightarrow X^+$. So, to obtain a monotonic and compactified coordinate $\tilde{u}$, we have to choose $c_u = 0$ and the plus sign in front of the $\arctan$ function for $R>X^+$, and correspondingly $c_u =1$ and the minus sign for $X^-<R<X^+$. Similarly, the $u$ coordinate reaches minus infinity exactly at $R\rightarrow X^-$. Thus, we choose $c_u=2$ and the plus sign for $R<X^-$. The coordinate $v$ is already monotonic for the trajectory of the surface, so we take $c_v=0$ and the plus sign in this case. However, to cover the vacuum part enclosed by $X=X^-$ and $X=0$ (see Fig. \ref{diagram}), we have to choose $c_v=1$ and the minus sign. Of course, we have to appropriately match signs and constants whenever the surfaces $X^-$ or $X^+$ are crossed. As for $\kappa$, it is simply a dimensionless, positive constant that allows us to adjust the form of the diagram. The above construction is a simplified (but equally general) version of the procedure presented in Ref. \cite{Schindler:2018wbx}.

To extend Eq. \eqref{comp} to the interior, we follow the method developed in Ref. \cite{Bobula:2024}. We define
\begin{equation} \label{integrals}
        u_{\mathrm{int}} = \int_0^T \frac{1}{a(\tilde{T})} \mathrm{d}\tilde{T} -r \,,\qquad
         v_{\mathrm{int}} = u_{\mathrm{int}}  + 2 r \, ,
\end{equation}
as the double null coordinates for the interior line element \eqref{intmetric}. Then, the extension of \eqref{comp} is given by
\begin{equation} \label{nullextended}
    \begin{split}
      \tilde{u}(T,r) &=\pm  \arctan\left[\kappa \, u\left(T\left\{u_{\mathrm{int}}(T,r) \right\}\right)\right] /\pi+ c_u \, , \\
        \tilde{v}(T,r) &=\pm  \arctan\left[\kappa \, v\left(T\left\{v_{\mathrm{int}}(T,r) \right\}\right)\right] /\pi+ c_v \, ,
    \end{split}
\end{equation}
where $T\{u_{\mathrm{int}} \}$ and $T\{v_{\mathrm{int}}  \}$ are, respectively, the inverse of $u_{\mathrm{int}}(T,r=r_b)$ and $v_{\mathrm{int}}(T,r=r_b)$. With the above coordinates, we can plot for instance the trajectory $(\tilde{u}(T,r=0),\tilde{v}(T,r=0))$, as displayed in Fig. \ref{diagram}. We point out that, in the limit $T\rightarrow \infty$, the whole $r$-constant congruence of timelike geodesics, with $0\leq r \leq r_b$, terminate at the same point of the conformal diagram. Since the scale factor $a$ goes to zero in this limit, the integral in Eq. \eqref{integrals} then diverges. In other words, for all the geodesics $0\leq r \leq r_b$, we have 
\begin{equation}
\lim_{T\rightarrow \infty} u_{\mathrm{int}} = \infty , \quad \lim_{T\rightarrow \infty} v_{\mathrm{int}} = \infty \, .\end{equation} 
Therefore, the values of the double null coordinates \eqref{integrals}, or alternatively \eqref{nullextended}, coincide in all the congruence at $T\rightarrow \infty$. Once can also check this statement by explicitly computing the integral via a change of variables $a=a(T)$ and utilizing Eq. \eqref{Tofa}.

We refer to Ref. \cite{Bobula:2024} for the detailed construction and limitations of the presented method for coordinate system extensions.

For the conformal diagram in the right in Fig. 2, being viable for classical OS collapse scenario \cite{Oppenheimer:1939ue}, we constructed the global coordinate system $(\tilde{u}_\mathrm{cl},\tilde{v}_\mathrm{cl} )$ in an analogous manner as presented in the above paragraphs.

The conformal diagram in the left in Fig. \ref{diagram} represents the collapse of a dust ball forming a Hayward black hole. The surface of the collapsing ball, at $r=r_b$, starts from $X\rightarrow \infty$ and crosses the pair of horizons $X^+$ and $X^-$, where $u\rightarrow \infty$ and $u \rightarrow -\infty$, respectively. Finally, it reaches $X=0$, where $u\rightarrow \infty$ and $v\rightarrow \infty$, in an infinite affine time according to co-moving observers. This endpoint is also reached in the diagram by the horizon $X=X^-$. Similarly, the trajectory of the origin of the radial coordinate ($r=0$) in the FRW interior also terminates there (see the previous paragraphs for the demonstration). Interestingly, the only radial null geodesic that reaches that point, i.e. the final collapsing point, is also a $X=X^-$ horizon generator (see Fig. \ref{diagram}). The rest of radial null geodesics either leave the collapsing dust ball forever at some point or never enter it. We will show below that the geodesic in question reaches the final point in infinite affine time. 

We rewrite the metric \eqref{ext} using outgoing Eddington-Finkelstein (EF) coordinates defined by a condition coming from \eqref{doublenull}, namely $\mathrm{d}u = \mathrm{d}t - \mathrm{d}X /F(X)$. This leads to
\begin{equation}
    \mathrm{d}s^2_{\mathrm{ext}}= -F(X) \mathrm{d}u^2 - 2\mathrm{d}u\mathrm{d}X + X^2 \mathrm{d}\Omega^2  \, .
\end{equation}
Let the coordinate $u$ be a parameter of a family of radial null geodesics, so that the tangent vector in EF coordinates is given by $q^{\alpha}=(1,-F(X)/2)$. We have for the four-acceleration $q^\alpha \nabla_\alpha q^\beta=-(\mathrm{d}F/\mathrm{d}X) q^\beta /2 $. Let $\lambda^*$ be the affine parameter. The parameters are related by the differential equation (see Chapter 1.3 of \cite{Poisson:2009pwt} for a general discussion)
\begin{equation}
    \frac{\mathrm{d^2}\lambda^* }{\mathrm{d}u^2 } = -\frac{1}{2} \frac{\mathrm{d}F }{\mathrm{d} X} \frac{\mathrm{d}\lambda^* }{\mathrm{d}u} \, .
\end{equation}
The solution is
\begin{equation}
    \lambda^*(u) = -2 c_1 \left(\frac{\mathrm{d}F}{\mathrm{d} X}\right)^{-1} \mathrm{exp}\left( -\frac{1}{2} \frac{\mathrm{d}F}{\mathrm{d} X} u  \right) +c_2 \, ,
\end{equation}
where $c_1$ and $c_2$ are integration constants. For the $X=X^-$ horizon generator, we have $\mathrm{d}F/\mathrm{d} X = const. <0$. Thus, when the final collapsing point is reached, with $u\rightarrow \infty$, the affine parameter $\lambda^*$ diverges. We have shown that the unique radial null geodesic reaches the final collapsing point in infinite affine time. 

We also observe that the regular quantum gravity core, represented by the timelike line $X=0$, is causally disconnected from the dust content. The diagram may be analytically extended beyond $X=X^-$ at the top to reproduce infinitely many copies of the Hayward static geometry there. 

\section{Discussion} \label{SECdis}

In this work, starting with the Hayward black hole metric and imposing junction conditions on the surface of a collapsing dust ball, we have constructed the complete geometry of a modified OS collapse model, which includes corrections arising from Planck scales. The Hayward exterior is static, whereas the interior is described by a homogeneous cosmology for which we have deduced corrected Friedmann equations. The resulting spacetime is geodesically complete. A power-law contraction of the interior dust matter precedes a smooth transition to a de Sitter phase. Although here the time orientation is chosen so that the model describes gravitational collapse, the cosmological evolution admits also a time-reversed version, in which de Sitter inflation is followed by a a graceful exit to a power-law behavior. Qualitatively, this cosmological evolution resembles the one found in the Starobinsky model \cite{Starobinsky:1980te}. Moreover, cosmic inflation plays in the model an unexpected role, namely, it provides a (quantum) mechanism that decelerates the collapsing matter so that a Schwarzchild-like singularity is never formed. 

The conformal diagram of our modified OS collapse model is shown in the left in Fig. \ref{diagram}. This diagram provides the global causal structure of the model. Before computing the diagram, it was not obvious whether the collapsing matter might terminate at the crossing point of the $X^-$ horizon and the core, $X=0$. Remarkably, the whole congruence of $r$-constant timelike geodesics actually ends up there forming a caustic. However, the fact that such point is reached after an infinite amount of affine time of the free-falling observers may suggest that the associated divergence of the energy density $\rho \rightarrow \infty$ would be absent in a more realistic collapse scenario accounting for the back-reaction of the Hawking radiation. The diagram representing the evaporation of the Hayward black hole (see Fig. 5 in Ref. \cite{Hayward:2005gi}) indicates that the black hole lifetime is finite for free-falling observers, so that, in our model, the whole black hole mass could evaporate before $T\rightarrow \infty$. Similarly, the only radial null geodesic reaching the final collapsing point arrives there in infinite affine time. Surprisingly, an FRW dust cosmology governed by Eqs. \eqref{intmetric} and \eqref{friedman} is geodesically incomplete. That is, for a cosmological model beyond OS, i.e. $r \in [0, \infty)$, there will be infinitely many radial null geodesics reaching $a \to 0$ in finite affine time, which implies geodesic incompleteness \cite{Borde:2001nh}. In contrast, when the cosmological dust ball is surrounded by a Hayward black hole, geodesic completeness is preserved. The detailed analysis of the corresponding model without the black hole is beyond the scope of this work. Nevertheless, our model teaches us a lesson, namely, that inflationary cosmological spacetimes, where $r \in [0, r_b]$, might be past (or future) complete when the exterior vacua are described by smoothly joined black hole metrics. In this way, such models may circumvent the Borde-Guth-Vilenkin theorem \cite{Borde:2001nh}, which states that spacetimes of this type are incomplete when $r \in [0, \infty)$.

Besides, the diagram indicates that the presence of the inner horizon may probably lead to instabilities, known in the literature as "mass inflation" \cite{Poisson:1989zz}. In particular, the main reason argued for the appearance of instabilities is that light rays traveling in the vicinity of the inner horizon would experience an arbitrarily high blueshift. However, the overall situation again might be cured by taking into consideration the back-reaction of the Hawking quanta, as advocated by Hayward (since a black hole evaporates, an inner horizon is no more real than the event horizon \cite{Hayward:2005gi}). Preliminary investigations supporting this possibility were recently conducted in Ref. \cite{Bonanno:2022jjp}. Nonetheless, the construction of a conformal diagram that incorporates the consequences of Hawking radiation, at least effectively, is beyond the scope of this work. It would be interesting to see whether Fig. 5 in Ref. \cite{Hayward:2005gi} is significantly modified when one considers the dust collapse presented here, instead of Vaidya-like ingoing energy flux forming the black hole.

Our model finds motivation in the recently proved fact that the Hayward metric is a vacuum solution of two different effective theories that incorporate quantum gravity corrections, namely, Quasi-Topological gravity \cite{Bueno:2024dgm, Arciniega:2018tnn} and LQG-inspired LTB models \cite{Giesel:2024mps}. Remarkably, the time dynamics of the scale factor which we have derived (see Fig. \ref{3wyk}) are similar to the evolution displayed in Fig. 1 of Ref. \cite{Arciniega:2018tnn} and in Fig. 5 of Ref. \cite{Giesel:2024mps}. This further supports the compatibility of our OS collapse model with those effective approaches. We note that there are prior works on the formation of the Hayward black hole \cite{Shojai:2022pdq, Malafarina:2022oka}. However, in these studies, the Hayward black hole was treated as a non-vacuum solution, leading to critical differences in the physical aspects compared to the model we propose. In Ref. \cite{Shojai:2022pdq}, a model based on the Einstein Field Equations was derived, wherein the Hayward black hole forms due to the collapse of a polytropic star with non-zero pressure and matter content present both in the interior and the exterior of the star. This model requires exotic matter (negative pressure develops in the star's interior), and the underlying dynamics differ from those derived in our work for a dust ball interior, as described in Eq. \eqref{friedman}. The other study, Ref. \cite{Malafarina:2022oka}, is based on EFEs coupled to nonlinear electrodynamics, where the presence of the electromagnetic field permits the Hayward black hole as a solution. An effective dust collapse model is proposed in this work, where 'semiclassical corrections' are identified with contributions from nonlinear electrodynamics. The resulting modified Friedmann equation (see Eq. (31) in Ref. \cite{Malafarina:2022oka}) explicitly depends on the magnetic charge, which is not the case for Eq. \eqref{friedman} in our work. While the trajectories of the collapsing objects discussed in Refs. \cite{Shojai:2022pdq, Malafarina:2022oka} may be compatible with the geometry we derived, the nonsingular nature of the studied geometry is not evident from these references\footnote{A detailed analysis, including the rigorous extraction of the conformal diagram, computation of the Kretschmann scalar, investigation of the final collapsing 'point', and examination of the only radial null geodesic arriving there, is notably absent in Refs.\cite{Shojai:2022pdq, Malafarina:2022oka} for the collapse into a Hayward black hole.}. Indeed,  we believe these studies overlook or underrepresent the most striking feature being present in our model: the role of the cosmic inflation in the collapse setting. The inflation (with the graceful entrance to the de Sitter phase) decelerates the collapsing matter, thereby preventing the formation of a singularity.

The model we have derived presents several clear advantages with respect to other alternatives discussed in the literature. In this sense, it is worth emphasizing the relative simplicity of the cosmological dynamics governed by Eq. \eqref{friedman}, the non-singular nature of the resulting spacetime and, most importantly, the inflationary phase that may successfully describe early stages of our universe. In contrast, bouncing OS dust collapse models coming from standard LQG techniques \cite{BobulaLOOPS,Bobula:2023kbo,Lewandowski:2022zce, Giesel:2022rxi, Giesel:2023hys, Giesel:2023tsj, Giesel:2024mps, Kelly:2020lec,  Kelly:2020uwj, Husain:2021ojz, Husain:2022gwp, Fazzini:2023ova, Fazzini:2023scu,Cafaro:2024vrw, Cipriani:2024nhx, Han:2024rqb} are singular, because they admit Reissner-Nordstr\"om-like singularities in vacuo\footnote{Notice also that the careful analysis carried out in Refs. \cite{Fazzini:2023ova, Giesel:2023hys} shows the singular behavior of the models proposed in Refs. \cite{Kelly:2020lec, Kelly:2020uwj, Husain:2021ojz, Husain:2022gwp} and of their causal structures.}. On the other hand, going beyond OS to the case of inhomogeneous dust collapse, the standard LQG techniques lead to shell-crossing singularities \cite{Fazzini:2023ova}. On the contrary, the geometry of the model proposed in this work is smooth everywhere, unlike that in Refs. \cite{Han:2023wxg, Han:2024rqb}. Let us also point out that there exist similarities between the dynamics of the model and the quantum-inspired gravitational collapse considered in Ref. \cite{Bambi:2013caa}. A Taylor expansion of the right-hand side of Eq. \eqref{friedman} around $\rho=0$ yields an infinite tower of (quantum) corrections characterized by consecutive powers of $\rho$ and $l$. These corrections were truncated at $\rho^3$ in Ref. \cite{Bambi:2013caa} (see the discussion around Eq. (23) in that work). However, the cosmological interiors obtained in Ref. \cite{Bambi:2013caa} may lead to timelike singularities in the exterior, just like in the case of standard LQG models, because the corresponding modified Friedmann equations are similar (see Eq. (24) in Ref. \cite{Bambi:2013caa} and Eq. (2) in Ref. \cite{Lewandowski:2022zce}). The inhomogeneous variant \cite{Liu:2014kra} of Ref. \cite{Bambi:2013caa} is also affected by comparable problems.  Thus, we see that a fully detailed study of the causal structure is necessary to decide whether the spacetime is actually non-singular. Indeed, the analysis in this work shows the importance of a rigorous construction of the conformal diagram in collapse models. It would be interesting to see whether some crucial features (for example, singularities) of the models in Refs. \cite{Barragan:2009sq, Bambi:2013caa, Saini:2014qpa,Frolov:2015bta,Frolov:2015bia,Kiefer:2019csi, Mosani:2020ena, Gozdz:2022dsa,Shafiee:2022jfx, Kiefer:2023zxt, Alonso-Bardaji:2023qgu, Duque:2023syb, Barca:2023shv} remain hidden because their conformal diagrams have not been obtained. Interestingly, our dust collapse model shows also a remarkable resemblance with the model obtained in Ref. \cite{Bonanno:2023rzk} using ASG. More concretely, the scale factor and the exterior metric function behave qualitatively in the same way (see Figs. 1 and 2 in that reference), and the dust energy density also diverges. Thus, one can expect that the conformal diagram obtained here (but absent in Ref. \cite{Bonanno:2023rzk}) and all the consequences derived from it would apply also to that case. However, our model does not require a running Newton gravitational ``constant'', a phenomenon which has not been (at least so far) experimentally supported.

Summarizing, our model represents a clear advance, with important improvements compared to other proposals published in the literature inasmuch as it simultaneously displays the following good physical properties. (i) It does not need to introduce new degrees of freedom, or (exotic) matter fields additional to those of the case of dust collapse in GR, or a running gravitational constant. (ii) It has a relatively simple cosmological dynamics that may successfully describe the early stages of our universe (with graceful exit from exponential inflation). (iii) The resulting spacetime is nonsingular (this fact was verified by a detailed analysis of the causal structure).

Finally, the modified Friedmann equations that we have obtained provide a possible route to analyse observational effects in cosmology. Especially, it would be interesting to study if they determine the dynamics of primordial perturbations, which would serve as seeds for the Cosmic Microwave Background (CMB).

\section*{Acknowledgements}
The author would like to thank Tomasz Paw{\l}owski and Guillermo A. Mena Marug\'{a}n for discussions and help in the preparation of the manuscript. This work was supported in part by the Polish National Center for Science (Narodowe Centrum Nauki –
NCN) grant OPUS 2020/37/B/ST2/03604.

\appendix

\bibliography{sample}

\end{document}